\documentclass[11pt,preprintnumbers,aps,amssymb,nofootinbib,amsmath,superscriptaddress]
{revtex4}
\usepackage{epsfig,epsf}
\usepackage{bm} 
\usepackage{slashed}
\usepackage{color} 
\newcommand{\beq}{\begin{equation}}
\newcommand{\beql}[1]{\begin{equation}\label{#1}}
\newcommand{\eeq}{\end{equation}}
\def\bal#1\gal{\begin{align}#1\end{align}}
\newcommand{\ball}[1]{\bal\label{#1}}
\newcommand{\eq}[1]{(\ref{#1})}

\newcounter{topiccounter}
\renewcommand{\b}[1]{{\bm #1}} 

\begin{document}

\title{Spontaneous topological transitions of electromagnetic fields in spatially  inhomogeneous  CP-odd domains}

\author{Kirill Tuchin}

\affiliation{Department of Physics and Astronomy, Iowa State University, Ames, Iowa, 50011, USA}

\date{\today}


\begin{abstract}

Metastable CP-odd domains of the hot QCD matter are coupled to QED via the chiral anomaly. The topology of electromagnetic field in these domains is characterized by  magnetic helicity. It is argued, using the Maxwell-Chern-Simons model, that spatial inhomogeneity of  the domains  induces spontaneous transitions of electromagnetic field between the  opposite magnetic helicity states.

\end{abstract}

\maketitle

\section{Introduction}\label{sec:a}

A possible existence of the metastable CP-odd domains in hot QCD matter has been actively discussed, especially in the context of the relativistic heavy-ion collisions \cite{Kharzeev:2009fn}. These domains  are described by a scalar field $\theta$ whose interaction with the electromagnetic field  $F^{\mu\nu}$ is given by the anomalous term in the QED Lagrangian \cite{Adler:1969gk,Bell:1969ts,Wilczek:1987mv,Carroll:1989vb,Sikivie:1984yz}
\ball{d3}
\mathcal{L}_A= -\frac{c_A}{4}\theta\tilde F_{\mu\nu}F^{\mu\nu}\,,
\gal
where   $\tilde F_{\mu\nu}= \frac{1}{2}\epsilon_{\mu\nu\lambda\rho} F^{\lambda\rho}$ is  the dual field tensor and 
\ball{d4} 
c_A= \frac{N_ce^2}{2\pi}\sum_f q_f^2 
\gal
is a constant. Together with the Maxwell's term $-(1/4)F_{\mu\nu}^2$, Eqs.~\eq{d3},\eq{d4} constitute the Maxwell-Chern-Simons (MCS) model \cite{Wilczek:1987mv,Carroll:1989vb,Sikivie:1984yz,Kharzeev:2009fn}, which is a useful tool for systematic study of the CP-odd effects in QED. The anomalous term \eq{d3} induces a number of remarkable effects, some of which may have already been experimentally observed, see reviews  \cite{Kharzeev:2013ffa,Huang:2015oca}.  

As has been recently pointed out in \cite{Chernodub:2010ye,Hirono:2015rla} the electromagnetic field inside the CP-odd domains is described by Chandrasekhar-Kendall (CK) states \cite{CK,Biskamp} which are spherical waves with definite magnetic helicity. Magnetic helicity determines the topology of the CK state and is a topological invariant  proportional to the number of twisted and linked flux tubes. 

The $\theta$-field is usually modeled by a spatially homogenous time-dependent function $\theta(t)$, which is suitable to study the temporal evolution of  topological configurations of electromagnetic field in matter with chiral asymmetry \cite{Joyce:1997uy,Hirono:2015rla}. Indeed, in the presence of the chiral imbalance, magnetic helicity is not conserved, hence an initial CK state, as well as $\theta$, undergo non-trivial topological evolution \cite{Joyce:1997uy,Hirono:2015rla} that manifests itself in transitions of electromagnetic field between the states with different magnetic helicity.  A significant change of the $\theta$-field typically occurs over the time $\tau$ which is of the order of the inverse electrical conductivity of the QCD matter. The lattice calculations indicate that near the critical point $\tau\sim 36$~fm. \cite{Ding:2010ga,Aarts:2007wj,Aarts:2014nba}. At significantly shorter time intervals the time-dependence of $\theta$ can be neglected. Thus, $\theta$ can be approximated by a constant for processes that occur at distances  much shorter than the domain size $R$,  \cite{Tuchin:2014iua,Tuchin:2016tks}. 

The main subject of this paper is interaction of electromagnetic fields with a CP-odd domain at time intervals shorter than $\tau$ and distances of order $R$. Since the CK states are spatially extended configurations, they are sensitive to the spatial variations of $\theta$, in particular to the finite size of the domain. Interaction of the electromagnetic field with the spatial gradient of $\theta$ induces transitions between different  CK states. Calculation of the corresponding transition rates is the main subject of this paper.  The main result is given by \eq{c55}. It  shows that the the only possible transitions are  $\{ l\to l\pm 1, h\to h\}$ and  $\{ l\to l, h\to -h\}$ with $m\to m$ in both cases. In particular, the transition rate between the states of opposite magnetic helicity is described by \eq{c62}. It indicates a possibility of spontaneous generation of magnetic helicity independently of the presence of the chiral imbalance.\footnote{Another mechanism that does not require the initial chirality imbalance for the magnetic helicity generation was recently discussed in \cite{Hirono:2016jps}.}

\section{Transitions between the CK states}\label{sec:b}

\subsection{Maxwell-Chern-Simons model}

Integrating by parts and dropping the full derivative, the anomalous term \eq{d3} can be cast in a more convenient form
\ball{d7} 
 \mathcal{L}_A= \frac{c_A}{2}\epsilon_{\mu\nu\lambda\rho}\partial^\mu \theta A^\nu \partial^\lambda A^\rho\,.
\gal 
Under the gauge $A^0=0$ \eq{d7} can be written as
\ball{d9}
 \mathcal{L}_A= \frac{c_A}{2}\left[ \dot\theta \b A\cdot \b B - \b\nabla\theta\cdot (\b A\times \b E)\right]\,.
\gal
As the precise spatial distribution of $\theta$ is not known, it makes sense to reduce the complexity of the problem, while keeping its essential features, by assuming that $\theta$ is time-independent  and its gradient given by 
\ball{a1}
 \b\nabla\theta = \b P f(r)\,,
\gal
where  ``the chiral polarization" $\b P$ is a constant, and $f$ is a smooth dimensionless function of the radial coordinate $r$ obeying the boundary conditions $f(0)=1$ and $f(\infty) = 0$.

\subsection{Chandrasekhar-Kendall  states}

The CK photons are elementary excitations of the electromagnetic field that have energy $\omega$ orbital angular momentum $l$, its projection $m$ and magnetic helicity $h$. It satisfies the dispersion relation 
$\omega=k$, where $k$ is the magnitude of the momentum that does not have a definite direction in the CK state. In the radiation gauge $A_0=0$, $\b\nabla\cdot \b A=0$ the CK photons are described in the coordinate representation by the wave functions 
\ball{b11}
\b A_{klm}^h(\b r, t)= \frac{1}{\sqrt{2kR}}h k\b W^h_{klm}(\b r) e^{-ik t }\,,
\gal
where  $\b W^h_{klm}(\b r)$ are the eigenfunctions of the curl operator
\ball{b13}
\b \nabla\times \b W^h_{klm}(\b r) = h k \b W^h_{klm}(\b r)
\gal
obeying the orthogonality conditions 
\ball{b15}
\int \b W^{h'*}_{k'l'm'}(\b r)\cdot \b W^{h}_{klm}(\b r)d^3 r=  \frac{\pi}{k^2}\delta(k-k')\delta_{ll'}\delta_{mm'}\delta_{h h'}\,.
\gal
For a typical photon energy, the domain radius $R$ is so large, that  $kR\gg 1$. This allows one to treat  the CK photon energy spectrum as approximately continuous. The electromagnetic potential can be written as 
\ball{b17}
\b A(\b r, t)= \sum_{lmh}\int_0^\infty \frac{Rdk}{\pi\sqrt{2kR}}\left(h  k a_{klm}^h\b W_{klm}^h(\b r) e^{-ik t}+ h.c.\ \right)\,,
\gal
where $a_{klm}^h$ is the operator obeying the usual bosonic commutation relations 
\ball{b18}
\left[a_{k'l'm'}^{h'},(a_{klm}^h)^\dagger\right]=\frac{\pi}{R}\delta(k'-k)\delta_{ll'}\delta_{mm'}\delta_{hh'}\,,
\gal
etc.  It is convenient to choose the quantization axis $z$ as the direction of vector $\b P$ and define the spherical coordinates $\theta$ and $\phi$ with respect to it. The functions $\b W^{h}_{klm}(\b r)$ can be expressed in terms of the spherical harmonics and the  orbital angular momentum operator $\b L = -i(\b r\times \b \nabla)$ as \cite{Jackson:1998nia} 
\ball{b19}
\b W^h_{klm}(\b r) = \b T^h_{klm}(\b r)-ih \b P^h_{klm}(\b r)\,,
\gal
where
\ball{b21}
\b T^h_{klm}(\b r)= \frac{j_l(kr)}{\sqrt{l(l+1)}}\b L[Y_{lm}(\theta,\phi)]\,,\quad \b P^h_{klm}(\b r)= \frac{i}{k}\b \nabla \times \b T^h_{klm}(\b r)\,, \quad l\ge 1\,.
\gal
Although functions $\b T_{klm}$ and $\b P_{klm}$ also form a complete set on a unit sphere (at fixed $k$), they do not  have definite magnetic helicity.

\subsection{Transition rate}

The scattering matrix element describing the scattering of the CK photon off the $\theta$-field is 
\ball{c11}
 \langle k'l'm'h'|S_A| k lmh\rangle= \frac{c_A}{2}\int d^4 x \langle k'l'm'h'|\dot \theta \b A\cdot (\b \nabla \times \b A)+ \b \nabla\theta \cdot (\b A\times \dot {\b A})| k lmh\rangle\,,
\gal
where $| k lmh\rangle\neq | k' l'm'h'\rangle$. Substituting \eq{b17} into \eq{c11} and using \eq{b13}-\eq{b15} one derives 
\ball{c13}
\langle k'l'm'h'|S_A| k lmh\rangle= \frac{c_A}{2}\frac{1}{2R}\int d^4 x e^{i(k'-k)t}\frac{1}{\sqrt{kk'}}\left\{ \dot\theta(h kk'^2+h'k'k^2 )\b W\cdot \b W'^*\right. \nonumber \\
\left. + \b \nabla\theta \cdot (\b W\times \b W'^*)(ikk'^2 +ik'k^2)hh'
\right\}\,,
\gal
where a shorthand notation is used: $\b W^h_{klm}= \b W$, $\b W^{h'}_{k'l'm'}= \b W'$.  In the case  of the time-independent  domain of radius $R$ described by \eq{a1}, \eq{c13} simplifies
\bal
\langle k'l'm'h'|S_A| k lmh\rangle & = \pi c_A \delta(k'-k)\frac{k^2}{2R}\int_0^\infty dr r^2f(r) \int d\Omega\, 2i \b P \cdot (\b W\times \b W'^*)hh'
\label{c15}\\
&= \pi c_A \delta(k'-k)\frac{k^2}{2R}\int_0^\infty dr r^2f(r) \, 2ihh'\b P\cdot \b C \,,\label{c16}
\gal
where 
\ball{c18}
 \b C= \int  \b W\times \b W'^*\, d\Omega=\int (\b T\times \b T'^*+ hh'\b P\times \b P'^*- ih \b P\times \b T'^*+ih' \b T\times \b P'^*)d\Omega\,.
\gal
The eigenfunctions $\b W'$ in \eq{c16} and \eq{c18} are evaluated at $k'=k$. The first integral in  \eq{c18} is proportional to 
\ball{c28}
\int \epsilon_{ijk}( L_j Y_{lm})( L_k^*Y_{l'm'}^*)\,d\Omega&= \int \epsilon_{ijk}Y^*_{l'm'} L_kL_jY_{lm}\,d\Omega= \frac{1}{2}\int \epsilon_{ijk}Y^*_{l'm'} [L_k,L_j]Y_{lm}\,d\Omega\nonumber\\
&=-i \int Y^*_{l'm'}L_i Y_{lm}d\Omega\equiv -i\langle l'm'|L_i|lm\rangle\,,
\gal
where the commutator $[L_k,L_j]=i\epsilon_{kjs}L_s$ was used. The explicit expression for the matrix element of the angular momentum is
\ball{c30}
\langle l'm'|L_i|lm\rangle=  \delta_{ll'}\big( m\delta_{mm'}\b e_z  
 &+\sqrt{l(l+1)-m(m-1)}\delta_{m',m-1}\b e_+
 \nonumber\\
 &+\sqrt{l(l+1)-m(m+1)}\delta_{m',m+1}\b e_-\big)\,,
\gal
where $\b e_\pm = \frac{1}{2}(\b e_x\pm i \b e_y)$.
 Eq.~\eq{c28} implies that 
\ball{c32}
\int \b T\times \b T'^* d\Omega = -i\frac{j_l^2(kr)}{l(l+1)}\langle l'm'|\b L|lm\rangle\,.
\gal

 The second integral in \eq{c18} reads
\bal
\int \epsilon_{ijk} P_j P'^*_k \, d\Omega= \frac{1}{k^2}\int \epsilon_{ijk}(\b \nabla \times \b T)_j(\b \nabla \times \b T'^*)_k\,d\Omega= 
\frac{1}{k^2}\int \epsilon_{kjs}T_s'^*\nabla_j\nabla_i T_k\,d\Omega\,,\label{c34}\\
= -\frac{j_l(kr)j_{l'}(kr)}{\sqrt{l(l+1)}\sqrt{l'(l'+1)}}\frac{1}{k^2}\int \epsilon_{kjs}(L_s^* Y_{l'm'}^*) (p_jp_i L_k Y_{lm})\,d\Omega\,,\label{c35}
\gal
where $\b p = -i\b \nabla$. Integrating by parts and using $\b p \cdot \b L=0$ and $[L_i,p_j]=i\epsilon_{ijk}p_k$, the integral in \eq{c35} can be rendered as
\ball{c37}
-\int \epsilon_{kjs}(L_s^* Y_{l'm'}^*) (p_jp_i L_k Y_{lm})\,d\Omega= -i\int Y_{l'm'}^*p^2  L_i Y_{lm}\, d\Omega = -\frac{il(l+1)}{r^2}\langle l'm'|L_i|lm\rangle\,,
\gal 
where I used $p^2Y_{lm}= l(l+1)Y_{lm}/r^2$. Thus, 
\ball{c39}
\int \b P\times \b P'^*\, d\Omega= -\frac{i}{k^2r^2}[j_l(kr)]^2\langle l'm'|\b L|lm\rangle\,.
\gal
Actually, integration by parts in \eq{c35} yields another term proportional to 
$$\int_0^\infty \b\nabla [r^2 f(r)j_l(kr)j_{l'}(kr)]dr\,.$$
However, it vanishes due to the boundary conditions imposed on $f(r)$, see \eq{a1}.

Turning to the third and fourth terms in \eq{c18} one obtains  after integrating by parts and using the gauge condition $\b \nabla\cdot \b T=0$
\bal
\int (-ih \b P \times \b T'^* +i h' \b T\times \b P'^*)\, d\Omega=-\frac{h+h'}{k}\int T'^*_i\b\nabla T_i \, d\Omega\nonumber\\
=-i\frac{j_l(kr)j_{l'}(kr)}{\sqrt{l(l+1)}\sqrt{l'(l'+1)}}\frac{h+h'}{k}\langle l'm'|L_i\b p L_i|lm\rangle\,. \label{c41}
\gal 
 Collecting \eq{c18},\eq{c32},\eq{c39} and \eq{c41} yields
\ball{c44}
\b C =&-i\bigg[  j_l^2(kr) \langle l'm'|\b L|lm\rangle \left( \frac{1}{l(l+1)} +\frac{hh'}{k^2r^2}\right) \nonumber \\
&+\frac{j_l(kr)j_{l'}(kr)}{\sqrt{l(l+1)}\sqrt{l'(l'+1)}}\frac{h+h'}{k}\langle l'm'|L_i\b p L_i|lm\rangle\bigg]\,.
\gal
The symmetry properties of the matrix elements imply that the first term in \eq{c44} describes transitions between the states with the same angular momentum $l\to l$, while the second one between the states with angular momentum different by one unit $l\to l\pm 1$. 

Plugging \eq{c44} into \eq{c16} and bearing in mind that $h^2=h'^2=1$ one derives
\bal
\langle k'l'm'h'|S_A| k lmh\rangle
&= \pi c_A \delta(k'-k)\frac{k^2}{R}\int_0^\infty dr r^2f(r) \nonumber\\
&\times\bigg\{ j_l^2(kr)\bigg[\b P\cdot \langle l'm'|\b L|lm\rangle\left( \frac{hh'}{l(l+1)}+\frac{1}{k^2r^2}\right)\bigg]\delta_{ll'}\nonumber\\
&+ \frac{h+h'}{k}\b P \cdot \langle l'm'|L_i\b p L_i|lm\rangle\frac{j_l(kr)j_{l'}(kr)}{\sqrt{l(l+1)}\sqrt{l'(l'+1)}}\bigg\} \,.
\label{c46}
\gal
The explicit expression for the matrix element   $\langle l'm'|L_i p_z L_i|lm\rangle$ is given by \eq{ap16} in the Appendix (recall that $\b P = P\b e_z$).  It is helpful to define the auxiliary functions
\bal
\mathcal{R}_{ll'}(k)&= \int_0^\infty dr r^2 f(r) j_l(kr)j_{l'}(kr)\,,\label{c49}\\
\mathcal{R}'_{ll'}(k)&= \frac{l(l+1)}{k^2}\int_0^\infty dr  f(r) j_l(kr)j_{l'}(kr)\,,\label{c50}\\
\mathcal{R}''_{ll'}(k)&= \frac{1}{k}\int_0^\infty dr r f(r) j_l(kr)j_{l'}(kr)\,. \label{c51}
\gal
Using all these equation and \eq{c30} in \eq{c46} one gets
\bal
\langle k'l'm'h'|S_A| k lmh\rangle= \pi c_A \delta(k'-k)\frac{k^2}{R} \delta_{m'm}
\bigg\{ \frac{P m}{l(l+1)}\left[ hh' \mathcal{R}_{ll'}(k)+\mathcal{R}'_{ll'}(k)\right]\delta_{ll'}\nonumber\\
+ i(h+h')P \left[ a_{lm} \delta_{l',l-1}+ b_{lm}\delta_{l',l+1}\right]\mathcal{R}''_{ll'}(k)\bigg\}\,,
\label{c52}
\gal
where the coefficients $a_{lm}$ and $b_{lm}$ are given by \eq{ap18} and \eq{ap19}. It is convenient to separate the part of the matrix element that describes the magnetic helicity flip $h' = -h$. To this end  one can multiply the first term in the curly brackets by the identity $1=\delta_{hh'}+ (1-\delta_{hh'})$.
The term  diagonal in all quantum numbers  only contributes to the wave function renormalization and can be dropped at the leading order. 

The transition rate form the initial CK state with quantum numbers $l$, $m$, $h$ to the final CK state with quantum numbers $l'$, $m'$ and $h'$ is given by 
\bal
w(lmh\to l'm'h')&= \frac{1}{t}|\langle k'l'm'h'|S_A| k lmh\rangle|^2\frac{Rdk'}{\pi}\label{c54}\\
&=  \frac{c_A^2k^4}{2R} \delta_{m'm}\bigg\{ 4P^2(\mathcal{R}''_{ll'}(k))^2\delta_{h'h} \left[a_{lm}^2\delta_{l',l-1} +b^2_{lm}\delta_{l',l+1} \right] \nonumber\\
&+ \frac{P^2 m^2}{l^2(l+1)^2}\big[\mathcal{R}'_{ll'}(k)-\mathcal{R}_{ll'}(k)\big]^2(1-\delta_{hh'})\delta_{ll'}\bigg\} \,.
\label{c55}
\gal
where the identity $(h+h')^2= 4\delta_{hh'}$ was used and one of the  delta functions is replaced by $ t/2\pi$. In particular, the rate of spontaneous  magnetic helicity-flip $h\to -h $ is 
\ball{c58}
w_\text{flip}(lm\to l'm')=  \frac{c_A^2 k^4}{2R}  \frac{P^2 m^2}{l^2(l+1)^2} \big[\mathcal{R}'_{ll}(k)-\mathcal{R}_{ll}(k)\big]^2\delta_{l'l}\delta_{m'm}\,.
\gal
 It is not difficult to verify that since  $kR\gg 1$,  $\mathcal{R}_{ll}\gg \mathcal{R}'_{ll} $ with the main contribution to the integral over $r$  arising from the distances $1/k<r<R$. Since at large $kr$ the spherical Bessel function can be approximated as  $j_l(kr)\approx (kr)^{-1}\sin(kr-\pi l/2)$, one finds 
\ball{c60}
\mathcal{R}'_{ll}(k)-\mathcal{R}_{ll}(k)\approx -\frac{1}{2k^2}\int_0^\infty  f dr\,.
\gal
Thus, the helicity-flip transition  rate is given by 
\ball{c62}
w_\text{flip}=  \frac{1}{8R} \bigg( c_A \frac{P m}{l(l+1)}  \int_0^\infty f dr \bigg)^2\,.
\gal
It is proportional to the domain radius and is independent of the CK state energy. 

\subsection{Estimates}

To estimate the transition rate for the quark-gluon plasma produced in relativistic heavy-ion collisions, one needs to know the value of $P$. Alternatively, one can solve \eq{a1} to obtain 
\ball{e1}
\theta(r)= P\int_0^r f(r')dr'+\theta_0\,,
\gal
where $\theta(0)= \theta_0$ is the value of $\theta$ in the domain's center. From the requirement that $\theta$ vanishes as $r\to \infty$ it follows that 
\ball{e2}
\int_0^\infty f(r)dr=-\frac{\theta_0}{P}\,.
\gal
Eq.~\eq{c62} can now be written as 
\ball{e3}
w_\text{flip}=  \frac{1}{8R} \bigg( c_A \frac{ m}{l(l+1)} \theta_0 \bigg)^2\,.
\gal
In \cite{Kharzeev:2007tn} the magnitude of the charge separation effect in a typical heavy-ion collision with $l=4$ is reproduced with $\theta_0  \simeq \pi$ for $N_f=2$. The domain size can be roughly approximated by the the sphaleron size $R=0.4$~fm \cite{Moore:2010jd}.
Substituting these estimates into \eq{e3}, yields for $l=m$ the magnetic helicity flip rate 
\ball{e4}
w_\text{flip}\sim 0.7\cdot 10^{-4} \,\text{fm}^{-1}\,.
\gal

\section{Summary}\label{sec:e}

The main result of this paper is Eqs.~\eq{c54}-\eq{c62} that give the transition rate between two states of electromagnetic field characterized by quantum numbers $l,m,h$ due to the spatial inhomogeneity of the CP-odd domain.  Eq.~\eq{c62} indicates that the spatially inhomogeneous CP-odd domains induce spontaneous flip of magnetic helicity. Thus, initially chirally symmetric electromagnetic field  can spontaneously acquire magnetic helicity by means of interaction with hot QCD matter.  Another possible way of magnetic helicity generation without any initial chirality imbalance was recently proposed in \cite{Hirono:2016jps}. Phenomenological implications of these effects in heavy-ion collisions, cosmology and condensed matter physics deserve a dedicated study.

\acknowledgments
This work  was supported in part by the U.S. Department of Energy under Grant No.\ DE-FG02-87ER40371.

\appendix
\section{The matrix element $\langle l'm'|L_ip_z L_i|lm\rangle$}\label{appA}

%
%

It is advantageous to employ  the raising and lowering operators $L_\pm = L_x\pm iL_y$ that act on the angular momentum eigenstates $|lm\rangle $ as 
\ball{ap1}
L_\pm |lm\rangle= \sqrt{(l\mp m)(l\pm m+1)}|l,m\pm 1\rangle\,.
\gal
Using \eq{ap1} one can reduce the matrix element of $L_ip_zL_i$ to the matrix elements of momentum:
\bal
&\langle l'm'|L_i p_z L_i|lm\rangle = \langle l'm'|L_zp_zL_z |lm\rangle+\frac{1}{2}\langle l'm'|L_+p_zL_- |lm\rangle+\frac{1}{2}\langle l'm'|L_-p_zL_+ |lm\rangle \label{ap2}\\
& =mm' \langle l'm'|p_z |lm\rangle + \frac{1}{2} \sqrt{(l'+ m')(l'- m'+1)} \sqrt{(l+ m)(l- m+1)}\langle l',m'-1|p_z |l,m-1\rangle \nonumber\\
&+ \frac{1}{2} \sqrt{(l'- m')(l'+ m'+1)} \sqrt{(l-m)(l+m+1)}\langle l',m'+1|p_z |l,m+1\rangle\,. \label{ap3}
\gal
The matrix elements of the momentum operator can be written as 
\bal
&\langle l'm'|p_z |lm\rangle= -i\langle l'm'|\frac{\partial}{\partial z}|lm\rangle
=-\frac{i}{r}\langle l'm'|\sin^2\theta \frac{\partial}{\partial \cos\theta}|lm\rangle\label{ap5}\\
&= \frac{i}{r}(l+1)\sqrt{\frac{(l+m)(l-m)}{(2l+1)(2l-1)}}\delta_{mm'}\delta_{l',l-1}-\frac{i}{r}
l\sqrt{\frac{(l-m+1)(l+m+1)}{(2l+1)(2l+3)}}\delta_{mm'}\delta_{l',l+1}\,, \label{ap6}
\gal
where the following recursive relation for the associate Legendre polynomials was used, see {\bf 8.731} in \cite{RG}
\ball{ap7} 
\sin^2\theta\frac{d}{d\cos\theta}P^m_l(\cos\theta)= \frac{1}{2l+1}\left[ (l+1)(l+m)P^m_{l-1}(\cos\theta)-l(l-m+1)P^m_{l+1}(\cos\theta)\right]\,.
\gal
Substituting \eq{ap6} into \eq{ap3} yields
\bal
\langle l'm'|L_ip_z L_i|lm\rangle
& = \frac{i}{r}\delta_{m'm}\Big\{ (l-1)(l+1)^2\sqrt{ \frac{(l-m)(l+m)}{(2l+1)(2l-1)}}\delta_{l',l-1}\nonumber\\
+& 
l^2(l+2)\sqrt{ \frac{(l-m+1)(l+m+1)}{(2l+3)(2l+1)}}\delta_{l',l+1}
\Big\}\,.
\label{ap14}
\gal
It is convenient to introduce a shorthand notation 
\ball{ap16}
\frac{\langle l'm'|L_ip_z L_i|lm\rangle}{\sqrt{l(l+1)}\sqrt{l'(l'+1)}} =\frac{i}{r}\delta_{m'm}\left[ a_{lm} \delta_{l',l-1}+ b_{lm}\delta_{l',l+1}\right]\,,
\gal
where
\bal
a_{lm}& = \sqrt{ \frac{(l-m)(l+m)(l-1)(l+1)^3}{l^2(2l+1)(2l-1)}}\label{ap18}\\
b_{lm}&=\sqrt{ \frac{(l-m+1)(l+m+1)(l+2)l^3}{(2l+3)(2l+1)(l+1)^2}}\,.
\label{ap19}
\gal


\end{document}